\documentclass[%
 reprint,
superscriptaddress,
 amsmath,amssymb,
 aps,
]{revtex4-2}

\usepackage{graphicx}
\usepackage{dcolumn}
\usepackage{bm}
\usepackage{hyperref}
\usepackage[mathlines]{lineno}

\begin{document}

\preprint{APS/123-QED}

\title{CE$\nu$NS Search with Cryogenic Sapphire Detectors at MINER: Results from the TRIGA reactor data and Future Sensitivity at HFIR}\vfill



\author{D.~Mondal}
\affiliation{National Institute of Science Education and Research, 
An OCC of Homi Bhabha National Institute, Jatni-752050, Odisha, India}

\author{W. Baker}
\affiliation{Department of Physics and Astronomy, Texas A\&M University, 578 University Dr, College Station, 77840, TX, US}

\author{M. Chaudhuri}
\affiliation{National Institute of Science Education and Research, 
An OCC of Homi Bhabha National Institute, Jatni-752050, Odisha, India}
\author{J. B. Dent}
\affiliation{Department of Physics, Sam Houston State University, 1905 University Ave, Huntsville, TX, 77340, US}

\author{R. Dey}
\affiliation{National Institute of Science Education and Research, 
An OCC of Homi Bhabha National Institute, Jatni-752050, Odisha, India}

\author{B. Dutta}
\affiliation{Department of Physics and Astronomy, Texas A\&M University, 578 University Dr, College Station, 77840, TX, US}

\author{V. Iyer}
\affiliation{Department of Physics, University of Toronto, 27 King's College Cir, Toronto, ON M5S 1A7, Canada}

\author{A. Jastram}
\affiliation{Department of Mechanical Engineering, Texas A\&M University, 202 Spence St, College Station, 77840, TX, US}

\author{V. K. S. Kashyap}
\affiliation{National Institute of Science Education and Research, 
An OCC of Homi Bhabha National Institute, Jatni-752050, Odisha, India}

\author{A. Kubik}
\affiliation{SNOLAB, Creighton Mine No. 9, 1039 Regional Road 24, Sudbury, ON P3Y 1N2, Canada}

\author{K. Lang}
\affiliation{Department of Physics, The University of Texas at Austin, 2515 Speedway, Austin, TX, 78712, US}

\author{R. Mahapatra}
\affiliation{Department of Physics and Astronomy, Texas A\&M University, 578 University Dr, College Station, 77840, TX, US}

\author{S. Maludze}
\affiliation{Department of Physics and Astronomy, Texas A\&M University, 578 University Dr, College Station, 77840, TX, US}

\author{N. Mirabolfathi}
\affiliation{Department of Physics and Astronomy, Texas A\&M University, 578 University Dr, College Station, 77840, TX, US}

\author{M. Mirzakhani}
\affiliation{Department of Physics and Astronomy, Texas A\&M University, 578 University Dr, College Station, 77840, TX, US}

\author{B. Mohanty}
\email{bedanga@niser.ac.in}
\affiliation{National Institute of Science Education and Research, 
An OCC of Homi Bhabha National Institute, Jatni-752050, Odisha, India}

\author{H. Neog}
\affiliation{School of Physics \& Astronomy, University of Minnesota, Minneapolis, MN, 55455, US}

\author{J.~L.~Newstead}
\affiliation{ARC Centre of Excellence for Dark Matter Particle Physics, School of Physics, The University of Melbourne, Parkville VIC 3010, Australia}

\author{M. Platt}
\affiliation{Department of Physics and Astronomy, Texas A\&M University, 578 University Dr, College Station, 77840, TX, US}

\author{S. Sahoo}
\affiliation{Department of Physics and Astronomy, Texas A\&M University, 578 University Dr, College Station, 77840, TX, US}

\author{J. Sander}
\affiliation{Department of Physics, University of South Dakota, 414 E Clark St, Vermillion, SD, 57069, US}

\author{L. E. Strigari}
\affiliation{Department of Physics and Astronomy, Texas A\&M University, 578 University Dr, College Station, 77840, TX, US}

\author{J. Walker}
\affiliation{Department of Physics, Sam Houston State University, 1905 University Ave, Huntsville, TX, 77340, US}

\collaboration{MINER Collaboration}\noaffiliation

\date{\today}

\begin{abstract}
We report on a search for coherent elastic neutrino--nucleus scattering (CE$\nu$NS) using cryogenic sapphire (Al$_2$O$_3$) detectors deployed at the Mitchell Institute Neutrino Experiment at Reactor (MINER), located near the 1~MW$_\text{th}$ TRIGA research reactor at Texas A\&M University. The experiment operated with a primary detector mass of 72~g and achieved a baseline energy resolution of $\sim 40$~eV. Using exposures of 158~g-days (reactor-on) and 381~g-days (reactor-off), we performed a statistical background subtraction in the energy region of 0.25--3~keV.  A GEANT4 simulation has been performed to understand the reactor-correlated background present in the data and it agrees with our observations. The resulting best-fit ratio of the observed CE$\nu$NS rate to the Standard Model prediction after rejecting the reactor induced background from the data with the help of simulation, is $\rho = 0.26\pm 1534.74~\mathrm{(stat)} \pm 0.05~\mathrm{(sys)}$ with a significance of $0.007 \pm 0.022~\mathrm{(stat)} \pm 0.001~\mathrm{(sys)}$. This low significance indicates a high background rate at low energies. To have enhanced sensitivity, the MINER collaboration plans to relocate the experiment to the 85~MW$_\text{th}$ High Flux Isotope Reactor (HFIR) at Oak Ridge National Laboratory (ORNL). With improved shielding, increased detector mass, and higher antineutrino flux, the upgraded setup is projected to achieve a 3$\sigma$ CE$\nu$NS detection within 30~kg$\cdot$days of exposure.

\end{abstract}

\maketitle

\section{Introduction\label{intro}}
Neutrinos are among the most elusive particles in the Standard Model (SM), interacting only via the weak interaction. While processes such as inverse beta decay (IBD) and neutrino-electron scattering have been widely used for neutrino detection, an alternative channel known as coherent elastic neutrino-nucleus scattering (CE$\nu$NS) offers a compelling probe of both SM and Beyond Standard Model (BSM) physics.

CE$\nu$NS is a neutral-current process in which a neutrino scatters coherently off an entire nucleus:
\begin{equation}
\nu + A(Z,N) \rightarrow \nu + A(Z,N).    
\end{equation}

This process becomes significant when the inverse momentum transfer is comparable or larger than the nuclear radius, ensuring the contributions from individual nucleons add coherently. Under these conditions, the CE$\nu$NS cross section is enhanced and approximately scales as the square of the neutron number, \( \sigma \propto N^2 \), making it a highly sensitive probe of weak interactions at low energies.

This neutral current electroweak interaction was first predicted in 1974 \cite{freedman} with a Standard Model differential cross section~\cite{CEvNS_crossSection1}
\begin{equation}
   \frac{d\sigma}{dT}(E_\nu, T) = \frac{G_F^2 M}{4\pi} \left(1 - \frac{MT}{2E_\nu^2}\right) [F(Q^2)]^2 \mathcal{Q}_W^2,
    \label{eq:CEvNS_crossSection}
\end{equation}
where \( E_\nu \) is the neutrino energy, \( T \) is the nuclear recoil energy, \( M \) is the mass of the target nucleus, \( G_F \) is the Fermi coupling constant, \( F(Q^2) \) is the nuclear form factor, $Q^2=-q^2$, $q$ is the 4-momentum transfer to the target nucleus and \( \mathcal{Q}_W = N - (1 - 4\sin^2\theta_W)Z \) is the weak charge of the nucleus. Having a small value of \( \sin^2\theta_W \approx 0.23867 \pm 0.00016 \) \cite{weak_mixing_angle}, the CE$\nu$NS interaction is predominantly sensitive to the number of neutrons inside the nucleus. Therefore, a detector material with a large atomic mass makes it a suitable choice for CE$\nu$NS search experiments. In contrast, a heavy nucleus produces very low recoil, making its detection very challenging.

Despite being predicted more than five decades ago, CE$\nu$NS was first observed only in 2017 by the COHERENT Collaboration using pion decay-at-rest neutrinos at the Spallation Neutron Source~\cite{COHERENT}. This milestone opened new avenues in neutrino physics, including the study of non-standard neutrino interactions, precision measurements of the weak mixing angle~\cite{neutrino_magnetic_moment1}, searches for light mediators~\cite{light_mediator}, and investigations into neutrino electromagnetic properties such as magnetic moment~\cite{neutrino_magnetic_moment1, neutrino_magnetic_moment2} and charge radius~\cite{neutrino_charge}. Beyond its role in constraining new physics, CE$\nu$NS offers a unique tool for studying the nuclear structure itself. Since the CE$\nu$NS cross section is primarily sensitive to the weak charge of the nucleus, which is dominated by neutrons, precise CE$\nu$NS measurements can provide access to the neutron distribution and form factor~\cite{neutron_form_factor}.

Nuclear reactors are ideal sources of low-energy (up to $\sim$10 MeV) antineutrinos, fully within the coherent scattering regime. Unlike accelerator sources, reactor-based CE$\nu$NS searches benefit from a continuous and intense flux of \( \bar{\nu}_e \) with average energies well suited for maximal coherence~\cite{coherency}. Furthermore, reactor antineutrino detection through CE$\nu$NS can serve as a powerful tool for reactor monitoring~\cite{reactor_monitoring}, nuclear safeguards, and understanding the irreducible neutrino background often referred to as the “neutrino floor” in direct dark matter searches~\cite{neutrino_irreducible_bkg}.

Several reactor-based CE$\nu$NS experiments are currently operational or under development, including CONUS~\cite{COUNSplus}, TEXONO~\cite{TEXONOresult}, CONNIE~\cite{CONNIE}, NUCLEUS~\cite{NUCLEUS}, Ricochet~\cite{Ricochet, Ricochet_comissioning}, and $\nu$GEN~\cite{NUGEN}. In this context, the Mitchell Institute Neutrino Experiment at Reactor (MINER) deploys low-threshold cryogenic sapphire detectors at close proximity to a reactor core. In addition to CE$\nu$NS searches, the same experimental setup has recently been employed to search for axion-like particles, demonstrating the broader physics reach of the MINER~\cite{MINER_newDirection_axion, MINER_axion}.

This paper presents the results of CE$\nu$NS search using data collected at MINER during its operation at the Texas A\&M University TRIGA reactor. We describe the experimental setup, detector response, event selection, and statistical treatment of the data. We also outline the prospects for significantly improved sensitivity in the upcoming MINER@HFIR deployment at the High Flux Isotope Reactor (HFIR) at Oak Ridge National Laboratory.

\section{MINER experiment}
\subsection{Reactor}

The Mitchell Institute Neutrino Experiment at Reactor (MINER) is a reactor-based CE$\nu$NS experiment located at the Nuclear Science Center (NSC) at Texas A\&M University. It utilizes a research reactor, called TRIGA (Training, Research, Isotope production, General Atomics), with an optimal power of 1 MW$_\text{th}$ fueled with low-enriched uranium (LEU, 20\% $^{235}$U). This reactor produces a high flux of antineutrinos ($\bar{\nu}_e$) via beta decay of fission fragments. At a baseline of $\sim 4$ meter from the core, the $\bar{\nu}_e$ flux is approximately $10^{10}$ cm$^{-2}$ s$^{-1}$, estimated using the method described in~\cite{CONNIE_signal_model}.

\subsection{Detector Concept and Cryogenic Setup}
MINER employs cryogenic sapphire detectors to achieve the low recoil threshold required for CE$\nu$NS detection~\cite{sapphire_detector1}. The detectors operate at $\mathcal{O}(10~\text{mK})$, enabling phonon-based readout with excellent energy resolution and minimal thermal noise. Cooling is provided by a BlueFors dilution refrigerator with a base temperature of approximately 10 mK.

\begin{table}[ht]
    \centering
    \caption{Specifications of the sapphire detectors used in the MINER tower.\\}
    \renewcommand{\arraystretch}{1.1}
    \begin{tabular}{l|c|c|c}
        \hline
        \rule[0 pt]{0pt}{10pt}\textbf{Detector} & \textbf{Top} & \textbf{Primary} & \textbf{Bottom} \\
        \textbf{Parameter} & \textbf{Veto} & \textbf{(Signal)} & \textbf{Veto} \\[2 pt]
        \hline
        Material & Al$_2$O$_3$ & Al$_2$O$_3$ & Al$_2$O$_3$ \\
        Diameter (mm) & 76 & 76 & 76 \\
        Thickness (mm) & 10 & 4 & 10 \\
        Density (g/cm$^3$) & 3.98 & 3.98 & 3.98 \\
        Mass (g) & 180 & 72 & 180 \\
        \hline
    \end{tabular}
    \label{tab:detector_details}
\end{table}

\begin{figure*}
\begin{minipage}[t]{0.49\linewidth}
{\small (a)\hfill\phantom\ \\}
    \includegraphics[width=0.75\linewidth]{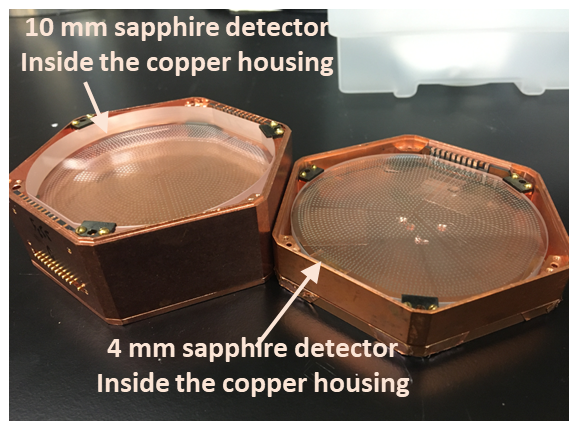}\\ \ \\ \ \\ 
    {\small(b)\hfill\phantom\ \\}
    \includegraphics[width=0.8\linewidth]{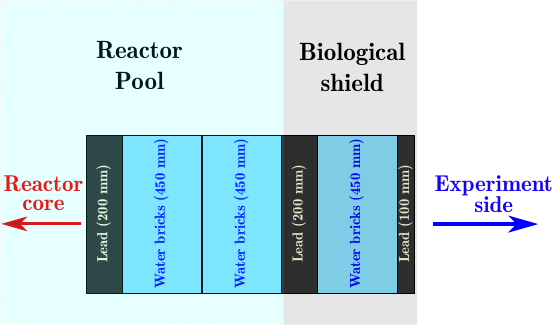}\\    
\end{minipage}
\begin{minipage}[t]{0.49\linewidth}
    {\small(c)\hfill\phantom\ \\}
    \includegraphics[width=\linewidth]{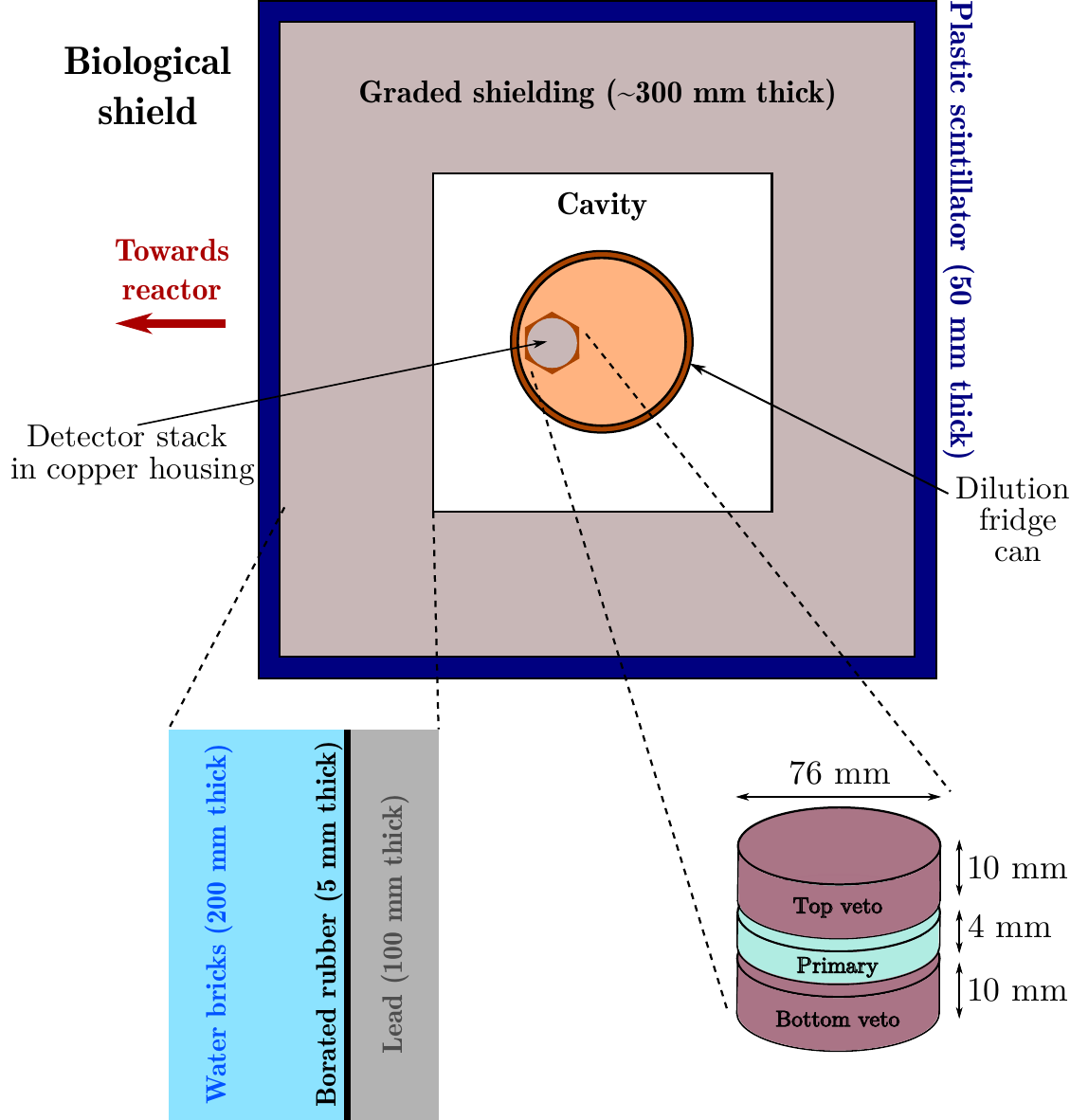}
\end{minipage}
    \caption{(a) Photograph of the primary (4 mm ) and veto (10 mm) sapphire detector with the phonon sensors photolithographically patterned on the surface. (b) Schematic of the cavity packed with layered shielding. (c) Detector shielding (top view) showing the placement of the detector stack inside the dilution fridge.}
    \label{fig:shielding}
\end{figure*}

During the data collection period, a tower of three sapphire detectors was installed inside the dilution refrigerator, which had a 4 mm thick primary detector~\cite{sapphire_detector} sandwiched between two 10 mm thick top and bottom active veto detectors. These veto detectors are used to reject backgrounds. Each crystal features photo-lithographically patterned phonon sensors on its top surface, known as quasiparticle-assisted electrothermal feedback transition-edge sensors (QETs)~\cite{QET}. These sensors detect athermal phonons produced during particle interactions. As phonons are absorbed, the TES transitions from superconducting to normal state, inducing a signal that is read out via a SQUID-based circuit~\cite{SQUID}. Each of the detectors is cylindrical, with four phonon channels: one outer annular channel A and three bulk channels B, C, and D as shown in Fig.~\ref{fig:shielding}(a).  The detectors were operated without a bias voltage. The specifications of the detectors are listed in Tab.~\ref{tab:detector_details}.

\subsection{Shielding and experimental configuration\label{sec:experimental_setup}}
The biological shield surrounding the reactor pool in the NSC facility is made of $\sim 2$ m thick high-density concrete ($\sim 3.5$ g/cm$^3$ density). It comprises a cavity inside it to facilitate near-reactor experiments as described in Ref.~\cite{MINER_bkg}. During the measurement period, the cavity was filled with layers of water and lead to stop gamma and neutrons produced inside the reactor core from reaching the experimental area. The cavity from the reactor core to the biological shield towards the experimental side was packed with layers of 200 mm thick lead, two 450 mm thick water brick layers, 200 mm thick lead layer, 450 mm thick water brick layer, and finally 100 mm thick lead as shown in Fig.~\ref{fig:shielding}(b). In addition to these, multi-layered shielding was also constructed around the dilution refrigerator containing detectors to suppress environmental and reactor backgrounds as shown in Fig.~\ref{fig:shielding}(c). From inside to outside, the detector shielding included a 100 mm thick lead, lined with a 5 mm thick layer of borated (52\%) rubber, a 200 mm thick water brick and a 50 mm thick active muon veto made from plastic scintillator. Water bricks were used to thermalise the neutrons, and borated rubber to capture the thermal neutrons. Additionally, the innermost lead layer stops the high-energy ambient gamma, as well as the gamma produced from the neutron capture in the borated rubber layer. The muon veto was not used for background rejection, as the low-energy excess background observed in Fig.~\ref{fig:sim_data}, which increases exponentially, could not be reduced by applying an anti-coincidence cut with the muon scintillator panels.

\subsection{Data Acquisition System}
The energy depositions and the event timestamps of each detector channel are digitized using a VME-based CAEN V1740D multichannel ADC. It offers 64 input channels, 12-bit resolution, and a sampling rate of 62.5 MS/s. Instead of hardware triggering, a software-based triggering algorithm (\texttt{SWT})~\cite{SWT_process} was employed to extract events from continuous voltage traces, enabling high-efficiency pulse selection in the low-energy regime.

\section{Measurement}
\subsection{Data Collection and Overview}
The analysis presented in this work is based on data collected at MINER during August - September 2022. This includes both reactor-on and reactor-off conditions, with exposures of 158 and 381 g-days, respectively. The reactor was operating in this duration with a power of $\sim1~\mathrm{MW_{th}}$. The datasets were segmented into sub-runs, each processed and calibrated independently before being combined for the final analysis. The reactor-off data were used to benchmark the background, capturing contributions from environmental radioactivity and cosmic rays. A statistical subtraction method is employed to eliminate the reactor off-background from the reactor on data. All reactor-on data were taken during the daytime operating hours of the reactor, whereas the reactor-off data were mostly collected at night. Therefore, there may be diurnal variations that were not taken into account in the background estimation.

\subsection{Event Reconstruction and Pulse Selection}
Continuous voltage traces from each detector channel were recorded in the data acquisition system and processed using the \texttt{SWT} to extract individual pulse events, each approximately 2 ms long. The extracted traces were further processed using the Optimal Filter (OF) algorithm to extract the energy information. OF algorithm estimates the pulse amplitude, enhancing the signal-to-noise ratio by performing a frequency domain fit between the observed pulse and a reference template.  This template is generated from the average of triggered signals, whereas the noise power spectral density (PSD) is derived from randomly triggered traces. The output amplitude of OF is proportional to the energy deposited in the detector. 

\begin{figure*}[t!]
    \centering
    \includegraphics[width=0.9\linewidth]{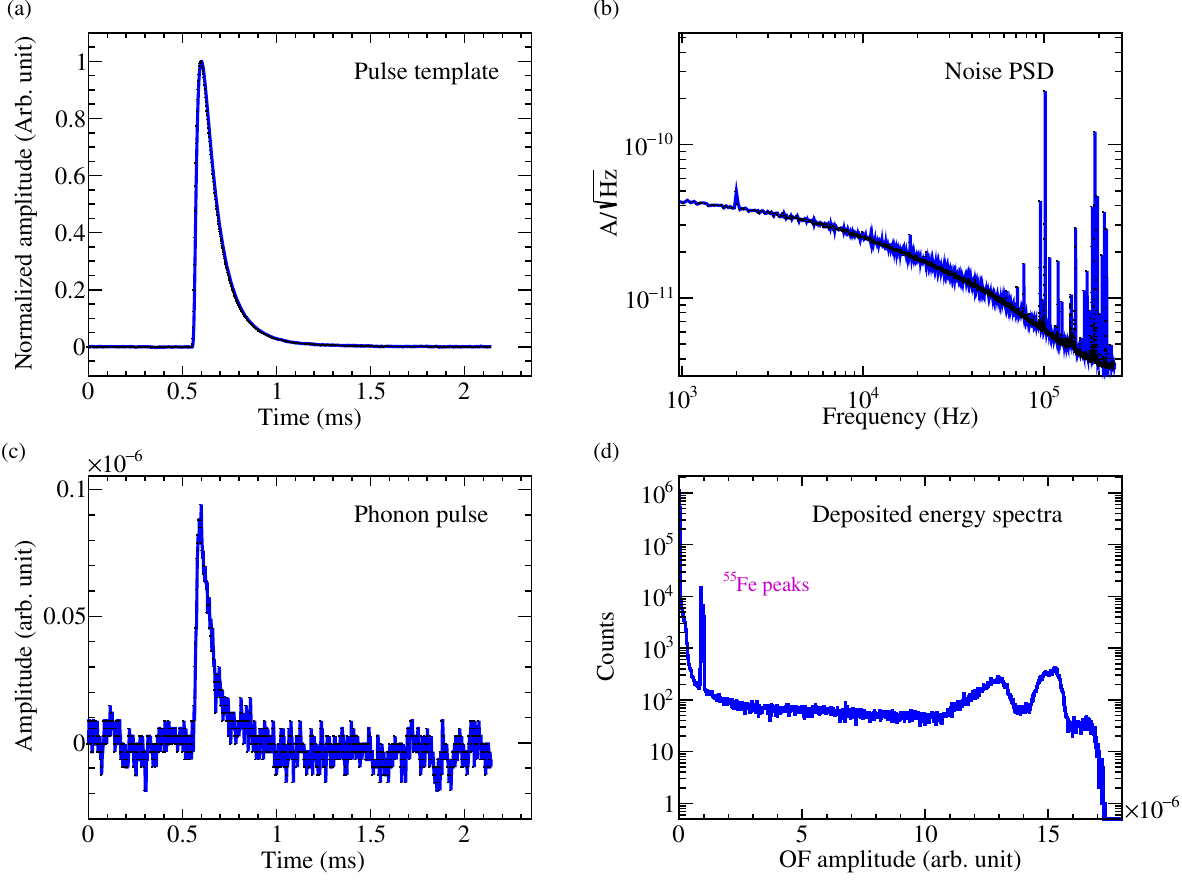}
    \caption{(a) A representative phonon pulse template, (b) noise performance, (c) a phonon pulse, and (d) the uncalibrated energy spectrum observed in the primary detector.}
    \label{fig:detector_performance}
\end{figure*}

A representative pulse template~\ref{fig:detector_performance}(a), noise PSD~\ref{fig:detector_performance}(b), and a typical triggered phonon pulse~\ref{fig:detector_performance}(c)  illustrate the components which go as input in OF. Figure~\ref{fig:detector_performance}(d) shows the resulting uncalibrated energy spectrum of the primary detector after summing up the OF amplitude of the four channels. The broad peak-like structures observed at higher amplitude in Fig.~\ref{fig:detector_performance}(d) correspond to high energy deposition in the detector, which leads to the TES saturation and incomplete estimation of the energy deposition. There are also some pile-up events and electronic glitches, which result in some bad pulses with large $\chi^2$ values. Events with poor fits (large $\chi^2$) are rejected to eliminate saturated, pile-up and instrumental artifacts.

\subsection{Energy Calibration}
\begin{figure}[h]
    \centering
    \includegraphics[width=0.9\linewidth]{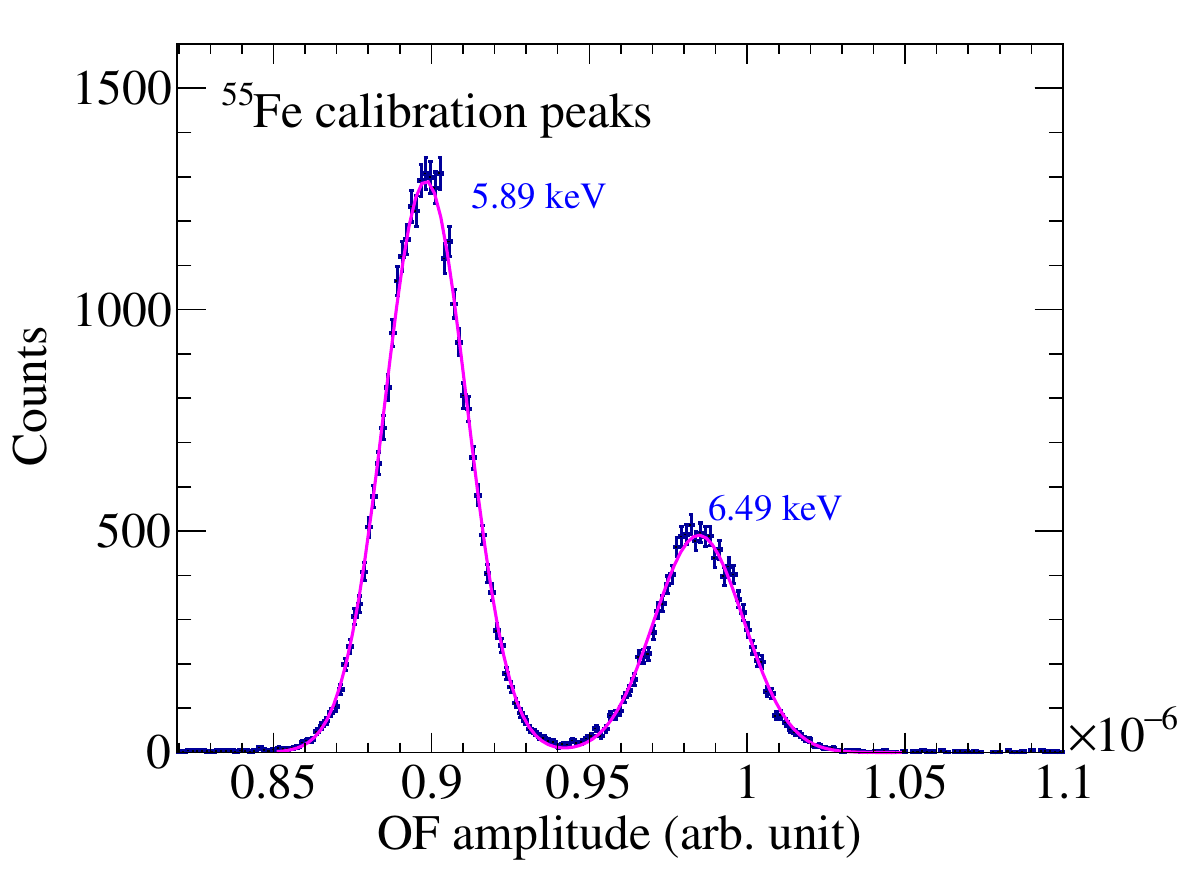}
    \caption{X-ray peaks at 5.89 keV and 6.49 keV from the $^{55}$Fe calibration source as observed in the primary detector. The double Gaussian fit function is shown in magenta colour to extract the mean positions.}
    \label{fig:detector_calibration}
\end{figure}

\begin{figure}[h]
    \centering
    \includegraphics[width=0.9\linewidth]{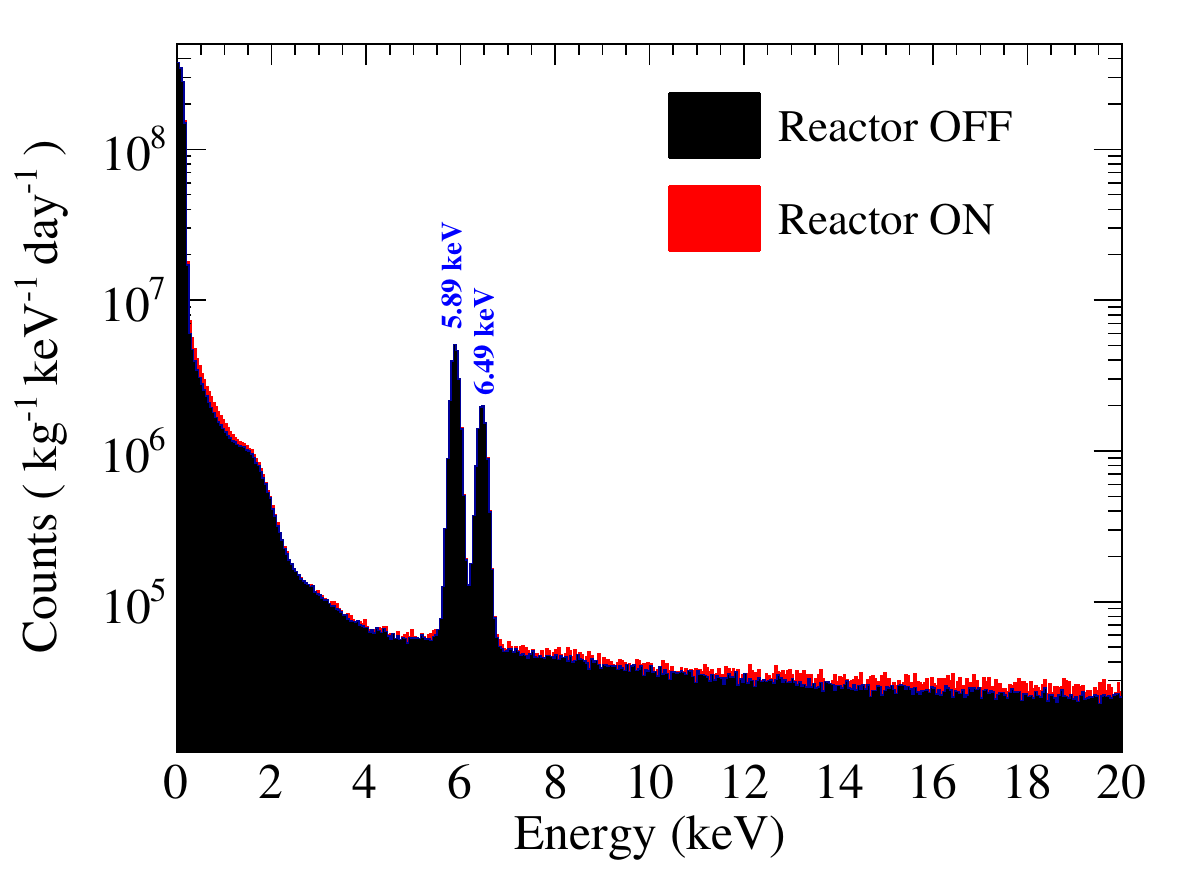}
    \caption{Energy spectra of events collected during 158 g-days (reactor ON) and 381 g-days (reactor OFF) within the 0–20 keV energy range. The two prominent peaks correspond to the 5.89 keV and 6.49 keV X-rays from the $^{55}$Fe calibration source.}
    \label{fig:reactor_on_off}
\end{figure}

The calibration of the three detectors was performed using two $^{55}$Fe sources. One source was placed between the top veto and the primary detector, and the second one was kept below the bottom veto detector, aligned with the central axis of the detector tower. These sources emit characteristic X-rays at 5.89 keV (K$_\alpha$) and 6.49 keV (K$_\beta$), which appear as well-resolved peaks in the response of the signal detector (Fig.~\ref{fig:detector_calibration}). The OF amplitude spectrum of each detector was calibrated by fitting the peaks with a double Gaussian and establishing a linear relationship between known energies and measured amplitudes. These calibration functions are then used to convert OF amplitudes in the full dataset into energy units (keV). The calibrated and normalized energy spectra of the on and off periods of the reactor are shown in Fig.~\ref {fig:reactor_on_off}.

The baseline energy resolution, denoted by $\sigma$, is defined as the standard deviation of the energy spectrum of randomly triggered noise traces. It quantifies the intrinsic noise of the detector and determines the minimum detectable energy. The range of resolutions across different sub-datasets for three detectors is listed in Table~\ref{tab:baseline_resolution}. The resolutions provided here correspond to the summed resolution of all four channels in the detector. Our achieved per channel resolution is 28 eV~\cite{sapphire_detector}, which corresponds to a sub-100 eV threshold for low energy events if the primary channel is used for energy estimates.

\begin{table}[h]
    \centering 
    \caption{Baseline energy resolution ($1\sigma$) for the three detectors across all sub-datasets.\\} 
    \renewcommand{\arraystretch}{1.1}
    \begin{tabular}{lcc} 
        \hline 
        \rule[0 pt]{0pt}{10pt}\textbf{Detector} & \textbf{Baseline resolution (eV)} \\[2 pt]
        \hline
        Top veto     &   107-125 \\
        Primary (Signal)   &  40-45 \\
        Bottom veto     &   110-142\\
        \hline
    \end{tabular}
    \label{tab:baseline_resolution} 
\end{table}

\subsection{Signal selection and background rejection}

Due to the low CE$\nu$NS cross section ($<10^{-40}$ cm$^2$), genuine signal events are expected to deposit energy only in the primary detector. This motivates the selection of only single scatter events, where the primary detector must register an energy deposition above the threshold ($>3\sigma$), while the top and bottom veto detectors remain within their noise levels ($<2\sigma$), defined by their baseline resolutions for each run. Additionally, the four phonon channels (A, B, C, D) in the primary detector must trigger within a 0.2 ms window, ensuring that the signal originates from a single correlated interaction. 

\begin{figure}[h]
    \centering
    \includegraphics[width=0.9\linewidth]{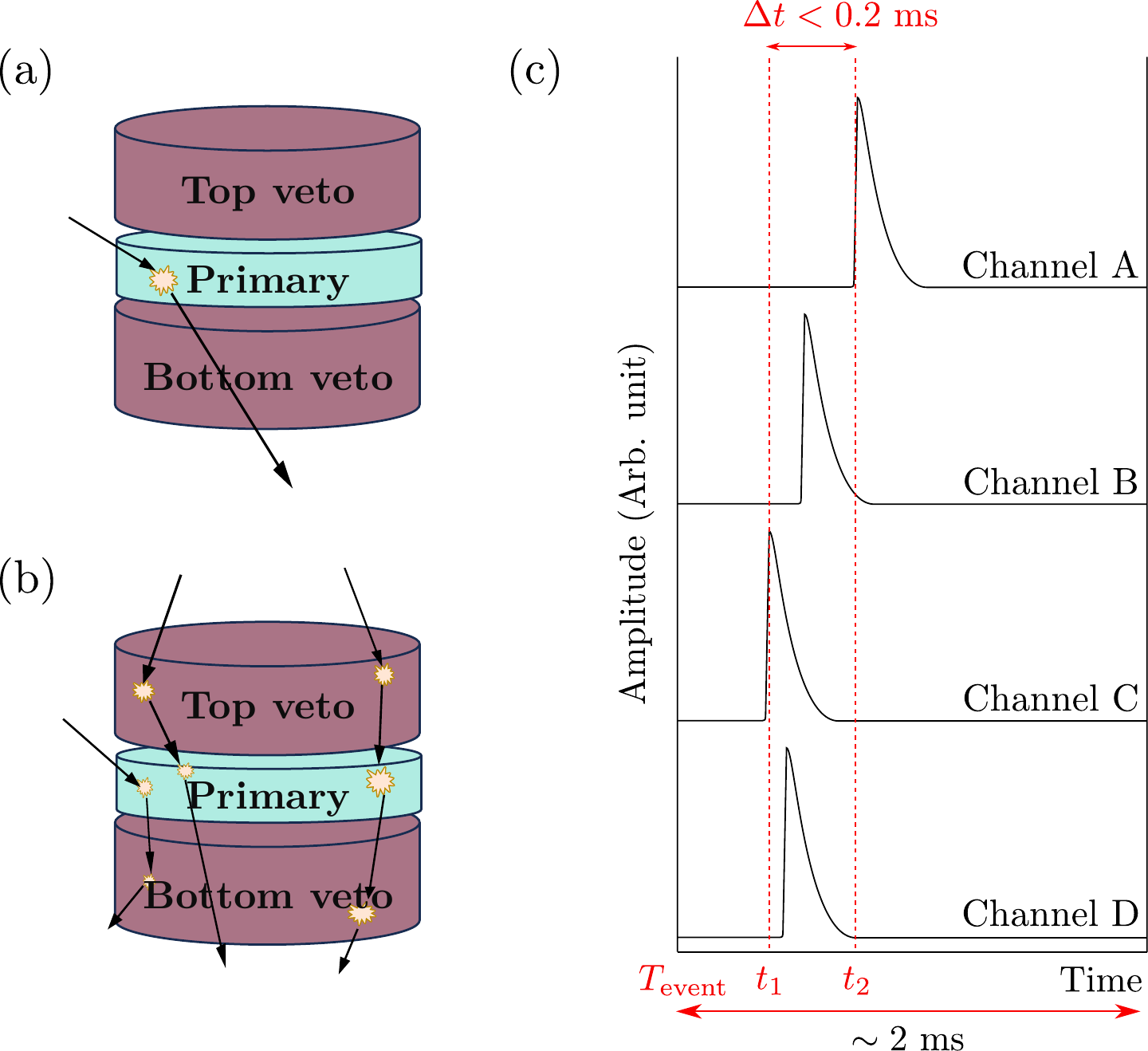}
    \caption{Schematic illustration of (a) a single scatter (signal-like), (b) a multiple scatter (background-like) event within the detector stack, and (c) time coincidence definition of the primary detector.}
    \label{fig:schematic_single_multipleScatter}
\end{figure}  

Moreover, energy depositions in channel A (outer channel) can lead to incomplete phonon collection, mimicking low-energy bulk events. These are identified using the A/sum parameter, which is the ratio of the OF amplitude in channel A to the sum of the OF amplitudes of all four channels.  A selection cut is applied to select events with a disproportionately low outer-channel contribution.

\begin{figure}[h]
    \centering
    \includegraphics[width=0.95\linewidth]{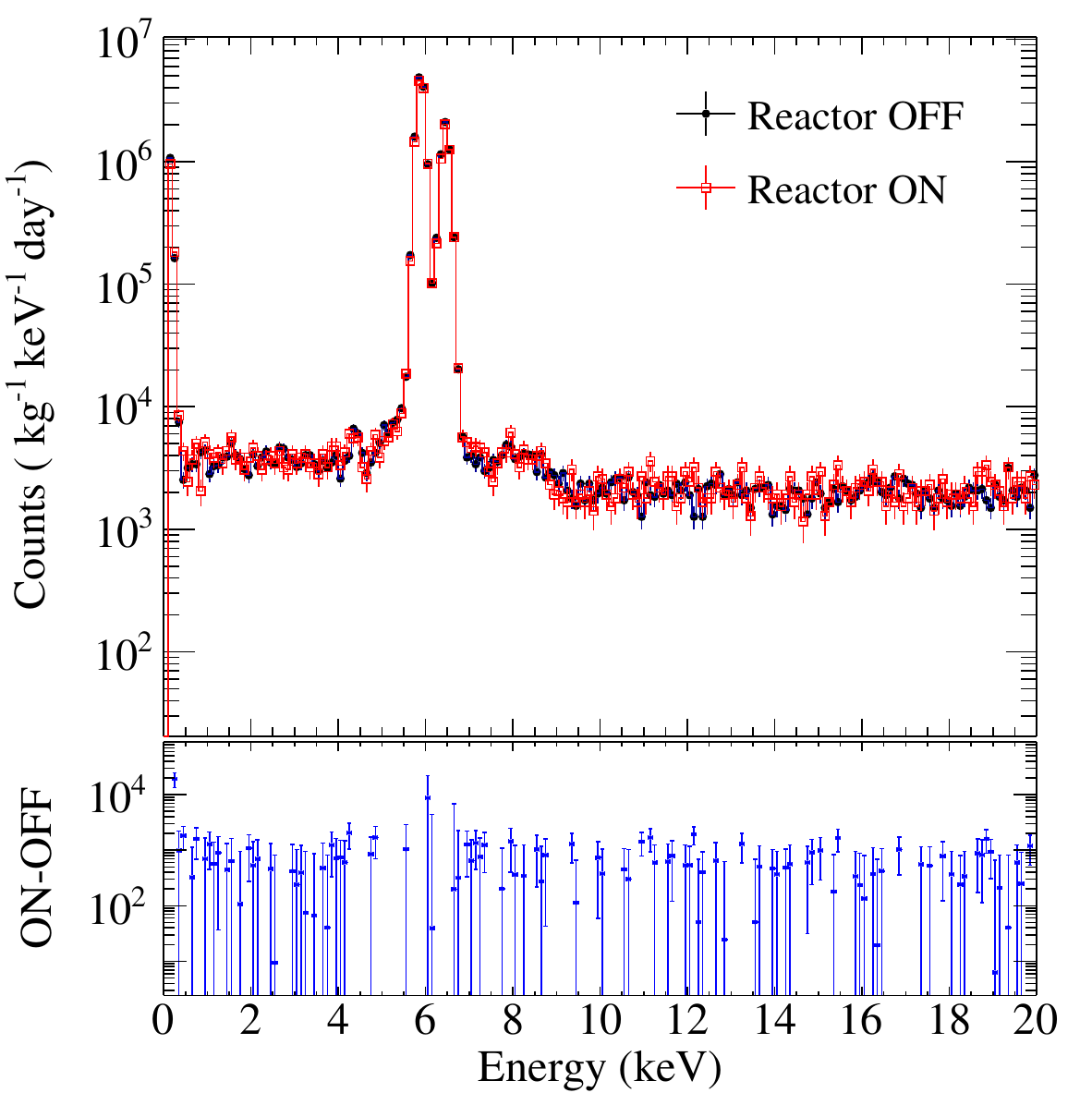}
    \caption{\textit{Top panel:} Event rate after applying the single scatter and fiducial volume cuts, shown per keV per kg per day in the 0–20 keV energy range. The two peaks correspond to the X-ray peaks of the $^{55}$Fe calibration source. \textit{Bottom panel:} Residual spectrum of signal-like events obtained by subtracting reactor-OFF from reactor-ON spectrum.}
    \label{fig:before_eff_corr}
\end{figure}

After all cuts, the differential event rate spectra for reactor-on and reactor-off periods are shown in Fig.~\ref{fig:before_eff_corr} (top panel). The residual spectrum obtained by subtracting reactor-off from reactor-on is shown in the bottom panel. This residual includes contributions from potential CE$\nu$NS events and additional reactor-induced gamma and neutron backgrounds.

\subsection{Efficiency correction and estimation of systematic uncertainties}
\begin{figure}[h]
    \centering
    \includegraphics[width=0.9\linewidth]{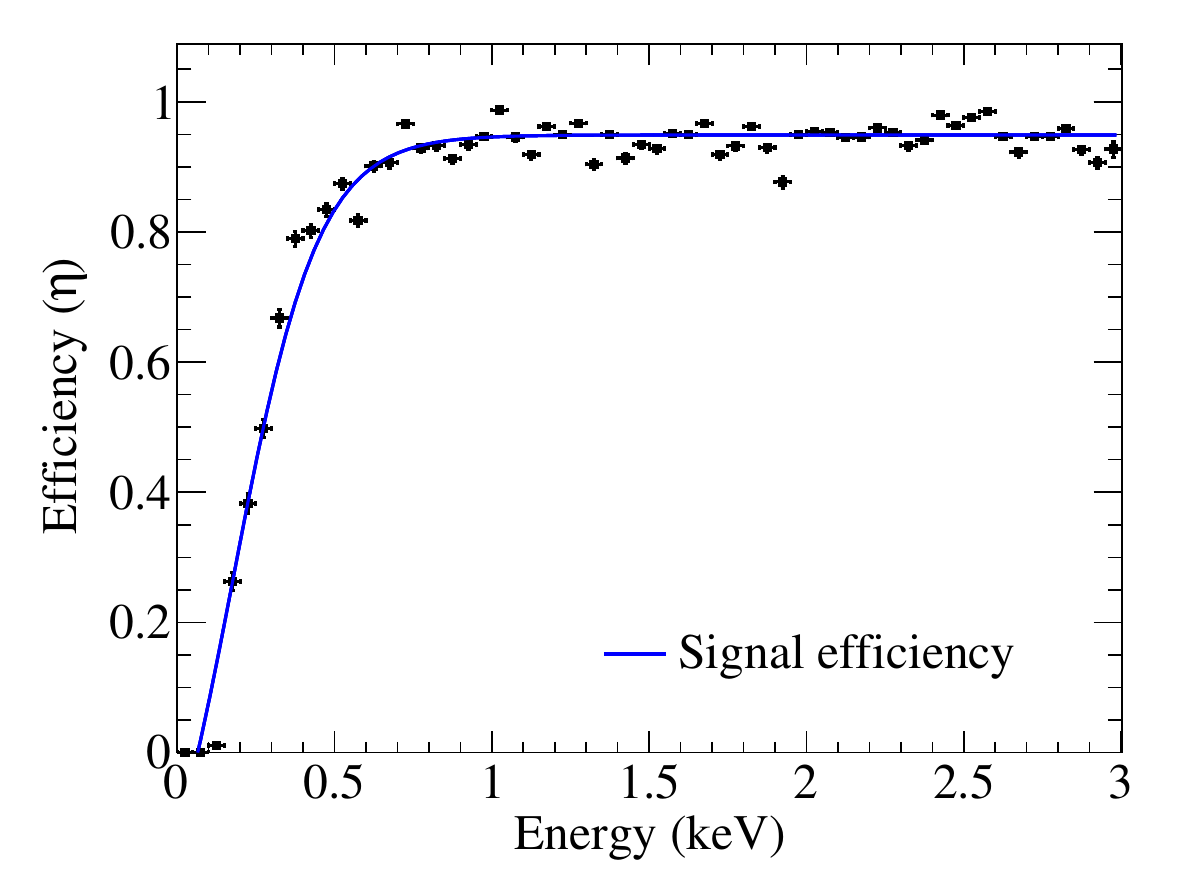}
    \caption{Signal efficiency resulting from the combined application of the trigger condition, $\chi^2$ cut, single scatter selection, and fiducial volume cut.}
    \label{fig:signal_efficiency}
\end{figure}

The application of various selection criteria, such as event trigger, $\chi^2$ cuts, single-scatter selection, and fiducialization can result in the loss of genuine signal events, especially at low recoil energies. To estimate this, a Monte Carlo simulation was performed in which artificial signal pulses were generated by injecting the average pulse template into random noise traces. The simulated dataset was binned in energy (0–3 keV range, 50 eV bins), with 100 fake events per bin. These were subjected to the same data processing and selection criteria as the real data. To quantify event selection efficiency, we also compute the passage fraction defined as the fraction of simulated events retained after each cut, and the result is presented in Tab.~\ref{tab:MC_passage}. The resulting energy-dependent efficiency is defined as the fraction of accepted events in each bin and is fitted using a sigmoid plus constant function as shown in Fig.~\ref{fig:signal_efficiency}. This function is later used to correct the measured event rate spectra.

\begin{table}[h]
    \centering
    \caption{Typical passage fractions of signal-like events after each selection cut separately.\\}
    \renewcommand{\arraystretch}{1.1}
    \begin{tabular}{l|c}       
        \hline
        \rule[0 pt]{0pt}{10pt}\textbf{Cut applied} & \textbf{Passage fraction (\%)} \\[2 pt]
        \hline
        $\chi^2$ Quality Cut       & $\sim$99 \\
        Single Scatter Requirement & $\sim$90 \\
        Fiducial Volume Cut (A/sum) & $\sim$91 \\
    \hline
    \end{tabular}
\label{tab:MC_passage}
\end{table}

The systematic uncertainties arising from the application of various analysis cuts were evaluated by varying selection thresholds and reanalyzing the datasets. The resulting uncertainties are propagated and summarized in Table~\ref{tab:systematic_combined}. Although the single scatter selection cut is the dominant systematic uncertainty (6.77\%) in this analysis, the statistical uncertainty from the measurement is much higher than the combined systematics. The experimentally measured integrated signal plus the background rate after incorporating the statistical and systematic uncertainties is $R_\mathrm{S+B} \approx 341 \pm 229~\mathrm{(stat)} \pm 23~\mathrm{(sys)}~\mathrm{kg}^{-1}\mathrm{day}^{-1}$.

\begin{table}[h!]
            \centering 
             \caption{Systematics of reactor on/off spectra related to different analysis cuts.\\} 
             \renewcommand{\arraystretch}{1.1}
            \begin{tabular}{l|c} 
                \hline 
                 \rule[0 pt]{0pt}{10pt}\textbf{Analysis cut} & \textbf{Systematic} \\ [2 pt]
                \hline
                $\chi^2$ quality cut & 0.055\% \\
                Single scatter  & 6.76\% \\
                Fiducial volume cut (A/sum)   & 0.27\%\\
                \hline
                All combined & 6.77\%\\
                
                \hline
            \end{tabular}
           
            \label{tab:systematic_combined} 
        \end{table}
%
%
\subsection{Geant4-based reactor on background simulation}

\begin{figure*}[htbp]
    \centering
    \includegraphics[width=\linewidth]{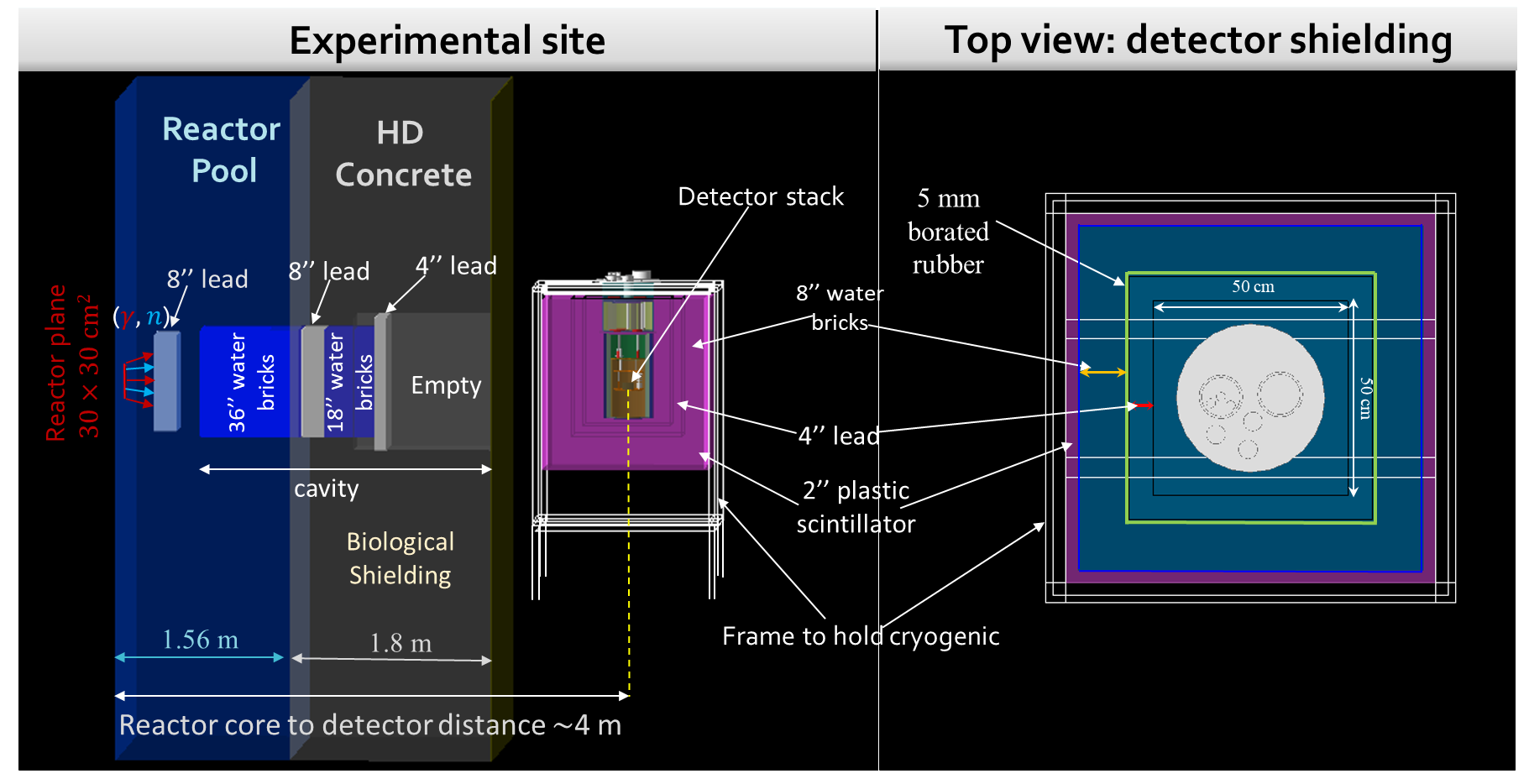}
    \caption{Geant4-based geometry model of the experimental site (side view) and the simulated composite shielding around the dilution refrigerator (top view). Different shielding materials are shown in different colors. Light gray: lead, Blue: water, Violet: plastic scintillator, Green: 52\%  borated rubber.}
    \label{fig:site_geometry}
\end{figure*}  

The complete experimental setup described in Section \ref{sec:experimental_setup} is modeled using the Geant4 (v11.2.1)\cite{geant4} Monte Carlo framework as shown in Fig. \ref{fig:site_geometry}. The motivation is to benchmark the experimental findings. Since this analysis is based on the statistical subtraction method, the on minus off reactor spectrum consists primarily of the CE$\nu$NS signal together with background contributions. Although small residual effects from cosmic and environmental sources may persist, the dominant background is anticipated to originate from reactor-induced gammas and neutrons, particularly fast neutrons.

\begin{figure}[htbp]
    \centering
    \includegraphics[width=0.9\linewidth]{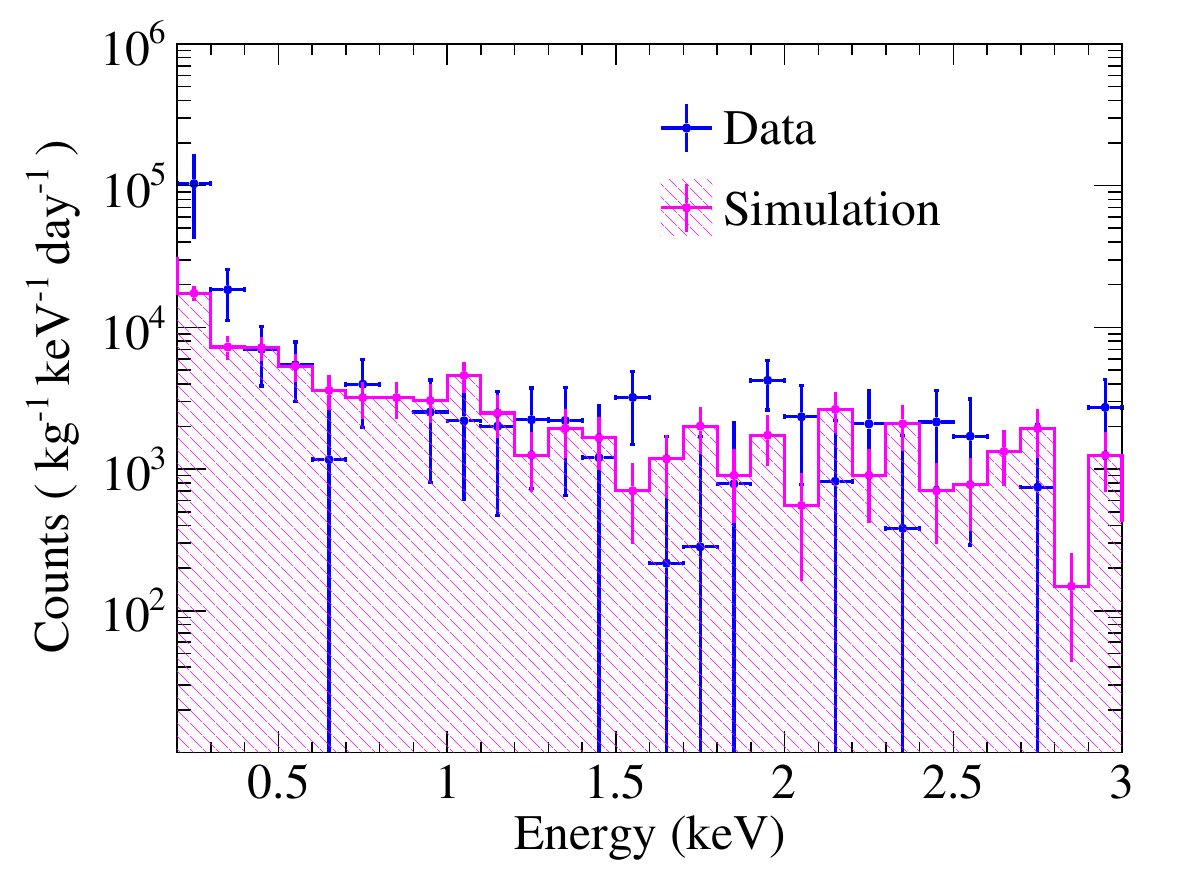}
    \caption{Comparison of the simulated single scatter spectra due to neutron and gamma backgrounds as estimated using the GEANT4 simulation and shielding design shown in Fig~\ref{fig:site_geometry} with the measured spectrum in the primary detector.}
    \label{fig:sim_data}
\end{figure}

 The ``Shielding" physics list is used to incorporate standard neutron and gamma interactions with matter. To reduce computational time and obtain good statistics, the simulated geometry is divided into two steps. In the first step, the reactor gammas and fast neutrons ($>100~\mathrm{keV}$) are generated using the MCNP energy spectra presented in \cite{MINER_bkg} from a plane $30\times30 ~\mathrm{cm}^2$ facing the experimental site from the reactor. The kinetic energy and direction of the particles reaching the end of the biological shielding are recorded. In the second step, particles (gamma and neutrons) are sampled from the recorded information and propagated towards the experimental shielding and eventually to the detector setup. An energy-dependent resolution, adapted from \cite{det_resolution}, is folded with the simulation. Figure~\ref{fig:sim_data} shows the comparison between the simulated single scatter spectrum and the measured single scatter spectrum in our detector. 

 The simulation spectrum reasonably agrees with the data except at low energies, where most low-threshold experiments observe low energy excess~\cite{Fuss:2022fxe}. From the simulation, we see that the spectrum obtained in the experiment is indeed dominated by reactor induced background, showing event rates of the same order. %

\section{Results and discussion}
\subsection{Signal modeling}

\begin{figure}[htbp]
    \centering
    \includegraphics[width=0.9\linewidth]{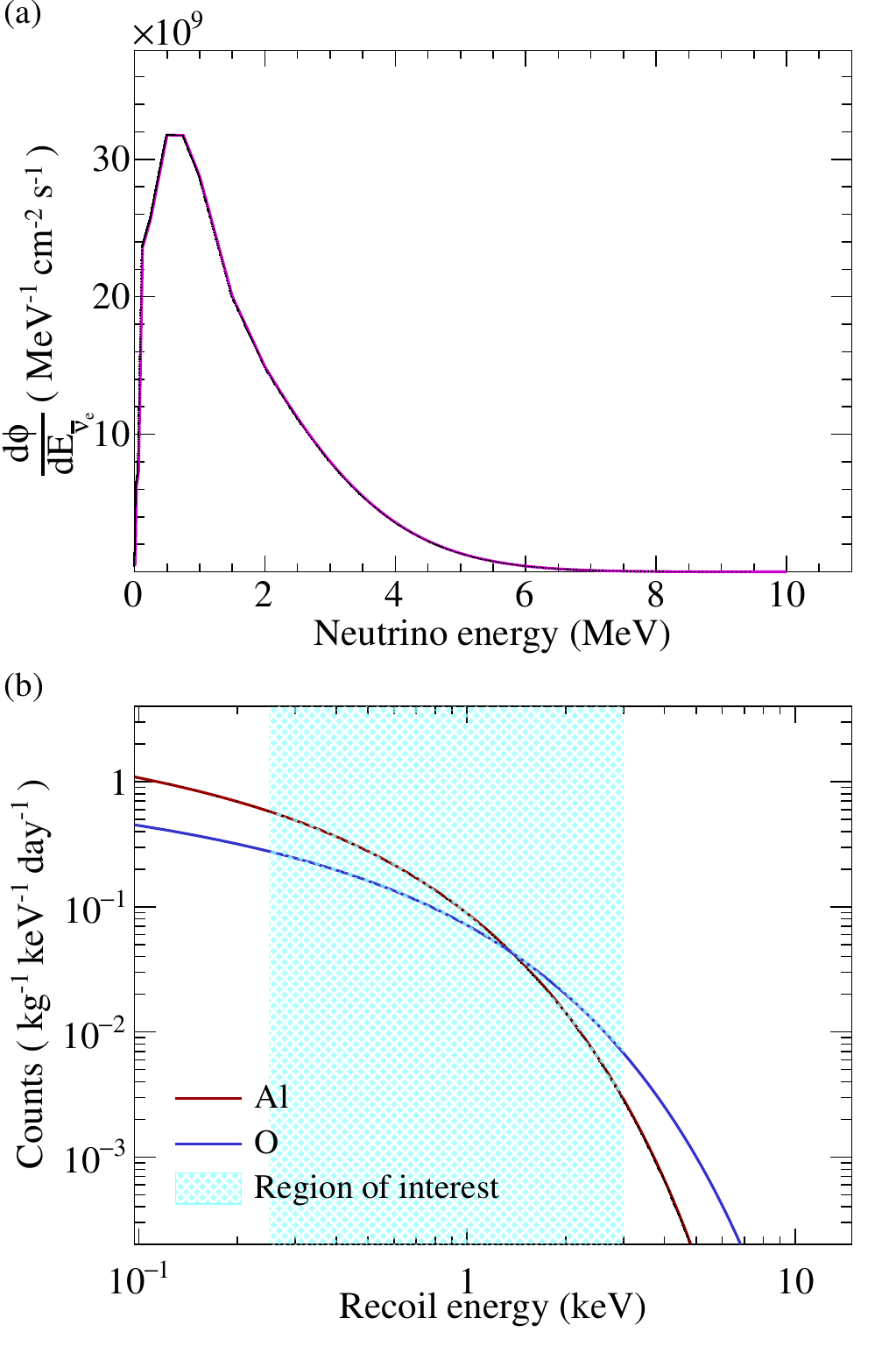}
    \caption{(a) Simulated $\bar{\nu}_e$ flux from the reactor (approximately, 1~MW$_\mathrm{th}$) incident on the detector located approximately 4 meters away from the core. (b) Expected CE$\nu$NS event rate spectra for interactions with Al and O nuclei in sapphire. The cyan-shaded region indicates the analysis region of interest (ROI) [0.25 - 3] keV.
}
    \label{fig:signal_modeling}
\end{figure}

A signal model based on the Standard Model (SM) prediction has been developed to estimate the expected CE$\nu$NS event rates in the current experimental configuration. The model uses the input parameters listed in Table~\ref{tab:signal_model parameters} and follows the formalism described in Ref.\cite{CONNIE_signal_model}.  The simulated differential electron antineutrino flux at the detector location, located $\sim4$ m from the reactor, is shown in Fig.~\ref{fig:signal_modeling}(a) as a function of neutrino energy. We have considered the contributions from the beta decay of four major fissioning isotopes: $^{235}\text{U},~ ^{238}\text{U},~^{239}\text{Pu},~^{241}\text{Pu}$. The neutron capture in $^{238}\text{U}$ produces $\bar \nu_e$ of energy $<1.3$ MeV (16\% of the total flux)~\cite{Kopeikin_neutron_capture}. These low energy neutrinos are not relevant to our analysis as only neutrinos with energy $>1.36$ MeV will contribute to the signal in our region of interest. The antineutrino spectrum above 2 MeV is well studied, with an uncertainty of 5\%, whereas the spectrum below is less understood. We have considered a conservative uncertainty of 30\%~\cite{flux_systematic} for the neutrino flux below 2 MeV. The $\bar\nu_e$ from the reactor having an energy greater than $\sim$1.36 MeV introduce an overall 16\% systematic uncertainty in the rate above the analysis threshold.

The corresponding CE$\nu$NS event rate spectra expected from individual interactions with aluminium and oxygen nuclei in the Al$_2$O$_3$ (sapphire) target are shown in Fig.~\ref{fig:signal_modeling}(b). The cyan-shaded band indicates the region of interest (ROI), defined from 0.25 to 3~keV nuclear recoil energy. The upper bound of 3~keV is chosen since the predicted signal rate drops below $10^{-2}$~kg$^{-1}$keV$^{-1}$day$^{-1}$ beyond this point, rendering the signal statistically insignificant.

For further statistical analysis, the signal rate is calculated as a weighted average of the spectra from both the aluminium and oxygen nuclei in differential rate units (DRU, per kg per keV per day), accounting for their stoichiometric ratios in the sapphire target.

\begin{table}[h]
            \centering 
            \caption{Parameters used for calculating the expected SM CE$\nu$NS signal.\\} 
            \renewcommand{\arraystretch}{1.1}
            \begin{tabular}{l|c} 
                \hline 
                \rule[0 pt]{0pt}{10pt}\textbf{Parameters} & \textbf{Values} \\[2 pt]
                \hline
                Reactor power & $\sim$1 MW$_{\mathrm{th}}$ \\
                Avg. fission fraction \cite{TRIGA_fission_fraction}& \textsuperscript{235}U (96.7\%), \textsuperscript{239}Pu (2\%), \\ &\textsuperscript{238}U (1.3 \%), \textsuperscript{241}Pu( $<$0.1\%) \\
                Baseline distance ($d$) & $\sim$4 m\\
                Duty cycle ($d_c$) & 100\% \\
                Target material &Al$_2$O$_3$ \\
                Detector efficiency ($\epsilon$)& 100\% \\
                Detector threshold & 250 eV \\
                Detector mass (m) & 72 g \\
                
                \hline
            \end{tabular}
            
            \label{tab:signal_model parameters} 
        \end{table}

\subsection{Statistical analysis\label{sec:statistical_analysis}}
A $\chi^2$ minimization analysis is performed in the region of interest (ROI), which spans $0.25-3$~keV, with an energy bin width of 50~eV. The test statistic is defined as:

\begin{equation}
    \chi^2(\rho)=\sum_i\left[\frac{N_i-\rho\times R_i^\mathrm{SM}}{\Sigma_i}\right]^2,
    \label{eq:chi2_analysis}
\end{equation}
where $N_i$ and $\Sigma_i$ denote the measured event rate and its corresponding uncertainty in the $i^{\mathrm{th}}$ energy bin, $R_i^{\mathrm{SM}}$ is the Standard Model (SM) predicted CE$\nu$NS event rate, and $\rho$ is a scaling factor representing the ratio of the measured to the predicted cross section. The 90\% confidence limit is extracted using a $\chi^2$ test statistics.
\begin{figure}[htbp]
    \centering
    \includegraphics[width=0.9\linewidth]{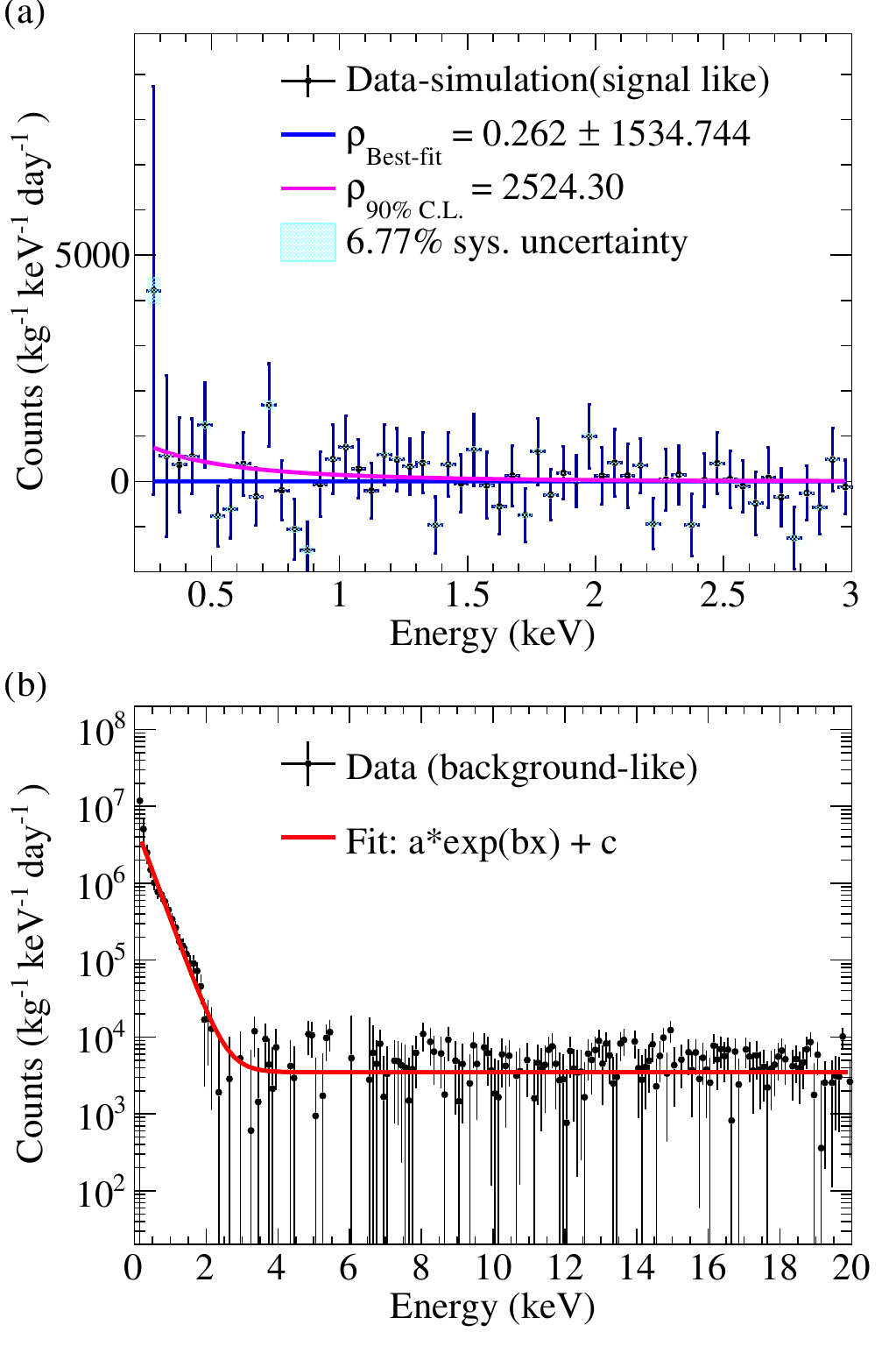}
    \caption{(a) Efficiency-corrected reactor ON--OFF signal spectrum after the subtraction of reactor ON background obtained from the GEANT4 simulation in the 250~eV to 3~keV energy range. The black and magenta lines represent the best-fit and 90\% confidence level (CL) upper-limit models, respectively. The cyan band indicates the systematic uncertainty. (b) Efficiency-corrected reactor ON--OFF background spectrum (multiple scatter events), fitted with an exponential plus constant model.}
    \label{fig:chi2_minimization}
\end{figure}  

 Since modeling the phonon response of the detector to the deposited energy is quite complex, the correction for fiducial volume in the Geant4 prediction was performed by scaling the single scatter spectrum by the ratio of events remaining after the fiducial volume cut to those after the single scatter cut, based on data. Figure~\ref{fig:chi2_minimization}(a) shows the efficiency-corrected ON--OFF residual spectrum after subtracting the reactor-correlated background obtained from simulation within the ROI. The spectra are overlaid with the best-fit and 90\% CL upper-limit curves derived from the $\chi^2$ analysis. The best-fit value of the scaling parameter is $\rho = 0.26\pm 1534.74~\mathrm{(stat)} \pm 0.05~\mathrm{(sys)}$, while the corresponding 90\% CL upper limit is $\rho = 2524.30~\pm ~403.89~(\text{sys})$. The rise in uncertainty near the 250~eV threshold reflects the reduced detection efficiency at the lowest energies.

Based on the SM prediction, the expected CE$\nu$NS event rate ($R_\mathrm{S}$) in the ROI during the reactor-on period is approximately $0.14~\mathrm{kg}^{-1}\mathrm{day}^{-1}$. In contrast, the experimentally measured integrated rate is $R_\mathrm{S+B} \approx 341 \pm 229~\mathrm{(stat)} \pm 23~\mathrm{(sys)}~\mathrm{kg}^{-1}\mathrm{day}^{-1}$. Whereas, the event rate after removing the reactor correlated background from the data is 59 $\pm$ 363~(stat) $\pm$ 4~(sys) $\mathrm{kg}^{-1}\mathrm{day}^{-1}$. The significance of the observed signal is quantified as:
\begin{equation}
    \alpha = \frac{N_\mathrm{S}}{\sqrt{N_\mathrm{S+B}}},
\end{equation}
where $N_\mathrm{S}$ and $N_\mathrm{S+B}$ are the total predicted signal and observed signal plus background events, respectively, obtained by multiplying the corresponding rates with the reactor-on exposure. The resulting significance is $\alpha = 0.007 \pm 0.022~\mathrm{(stat)} \pm 0.001~\mathrm{(sys)}$.

Notably, the reactor ON--OFF subtracted event rates for single-scatter (signal-like) events are of the same order of magnitude as those for multiple-scatter (background-like) events, further underscoring the dominant contribution of reactor-induced backgrounds in the region of interest. 
The observed enhancement in the event rate, at the level of $\mathcal{O}(10^4)$ above the SM prediction, coupled with the low statistical significance, indicates that the residual spectrum is dominated by reactor-correlated background contributions in the low-energy region. As shown in Fig.~\ref{fig:chi2_minimization}(b), the background spectrum exhibits an exponential plus constant behavior, with amplitudes in the range of $\mathcal{O}(10^4\text{ to }10^6~\mathrm{kg}^{-1}\mathrm{keV}^{-1}\mathrm{day}^{-1})$, confirming the substantial background presence in ROI.

\section{Concluding remarks and prospects }

The MINER collaboration successfully installed and operated a cryogenic detection setup near the $\sim$1~MW$_\mathrm{th}$ TRIGA research reactor at the Nuclear Science Center (NSC) for a search of coherent elastic neutrino--nucleus scattering (CE$\nu$NS). Data were collected during August–September 2022 with exposures of 158 and 381~g-days for reactor-on and reactor-off periods, respectively. The detectors used in this campaign were sapphire (Al$_2$O$_3$) crystals, with the primary 72~g detector achieving an excellent baseline energy resolution of $39.99 \pm 0.13$~eV.

\begin{figure}[h]
    \centering
    \includegraphics[width=0.9\linewidth]{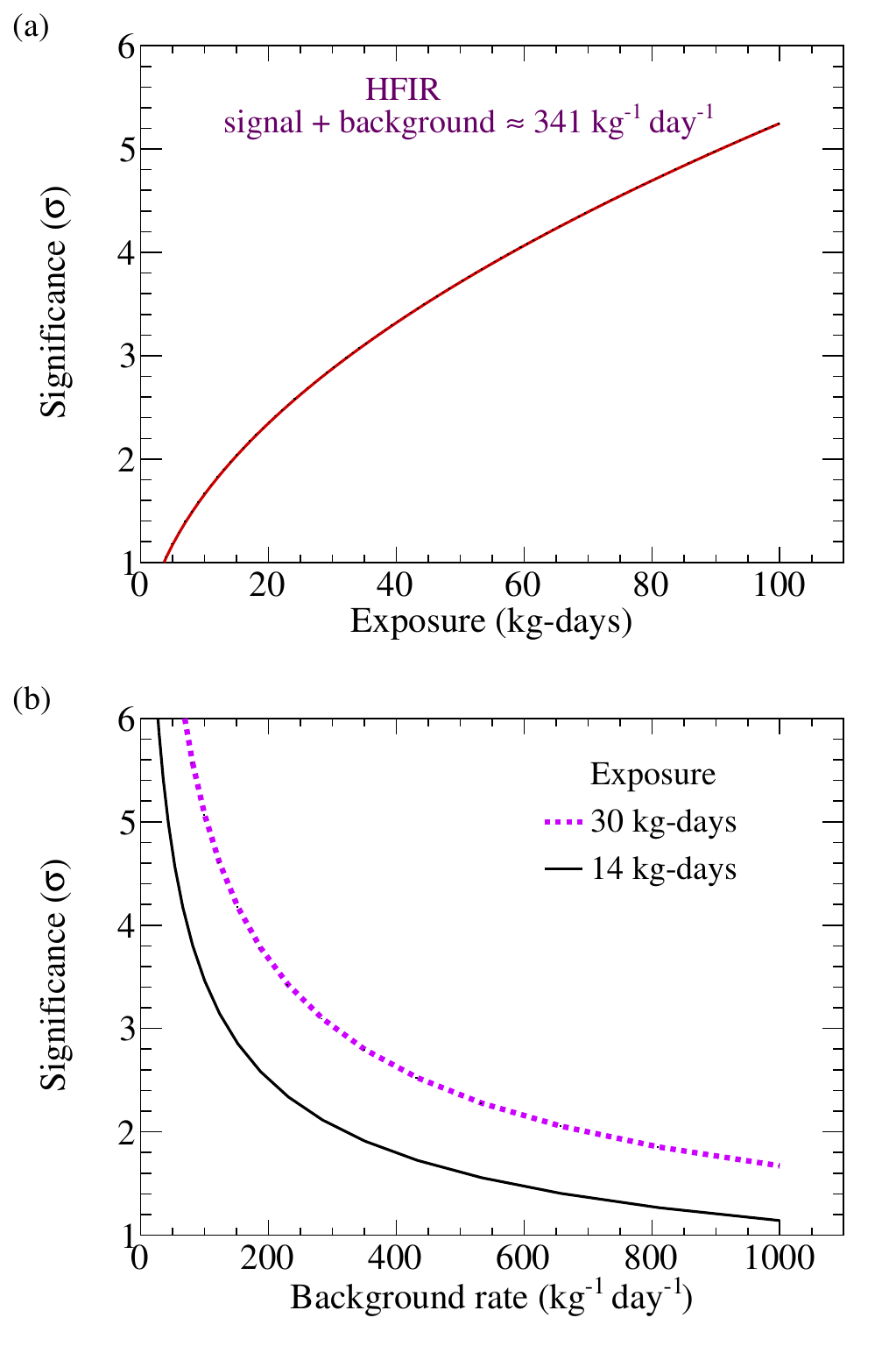}
    \caption{(a) Projected exposure (kg-days) required at the HFIR site to achieve 3$\sigma$ and 5$\sigma$ significance levels for CE$\nu$NS detection, assuming a signal+background rate of $\sim$341~kg$^{-1}$day$^{-1}$. (b) Required background levels to achieve 3$\sigma$ and 5$\sigma$ significance within 30 and 14~kg-day exposures, respectively.}
    \label{fig:HFIR_prospect}
\end{figure}

The analysis presented in this work evaluates the sensitivity of the current MINER configuration to CE$\nu$NS. The observed best-fit signal strength relative to the Standard Model prediction yields $\rho = 0.26\pm 1534.74~\mathrm{(stat)} \pm 0.05~\mathrm{(sys)}$. The  detection significance is $0.007 \pm 0.022~\mathrm{(stat)} \pm 0.001~\mathrm{(sys)}$, indicating that the observed excess is consistent with residual background contributions, as also supported by our simulation results. The data analysis employs rigorous pulse reconstruction and calibration procedures using $^{55}$Fe sources, enabling precise energy scale determination across detectors. A signal efficiency correction method, based on template injection into real noise traces has been developed. Systematic uncertainties from all analysis steps, including pulse shape quality cuts, single scatter selection, and fiducial volume definition, were carefully evaluated and propagated. These procedures ensure a proper sensitivity estimate despite challenging low-energy backgrounds.

Nevertheless, the low statistical significance suggests that the signal region is dominated by background, which is further supported by the simulation results. Even with enhanced shielding, the current event rate ($\sim 0.14$~kg$^{-1}$day$^{-1}$) remains insufficient to reach competitive sensitivity within a reasonable time frame using the present setup, even after assuming 100\% detection efficiency.

To address these limitations and significantly improve sensitivity, the collaboration plans to relocate the experiment to the 85~MW$_\mathrm{th}$ High Flux Isotope Reactor (HFIR) at Oak Ridge National Laboratory~\cite{reactor_HFIR2}. The HFIR facility provides a higher $\bar{\nu}_e$ flux, and the MINER setup will be optimized with improved compact shielding and an increased detector payload, featuring multiple primary detectors within the tower.

HFIR is a pressurized, light-water-cooled and -moderated reactor fueled with high-enriched uranium (HEU, 93\% $^{235}$U), currently operating at 85~MW$_\mathrm{th}$. The planned detector deployment will be located approximately 5 meters from the reactor core.

Sensitivity projections for MINER at HFIR are shown in Fig.~\ref{fig:HFIR_prospect}. With the current noise level and experimentally achieved signal plus background rate of $\sim 341$~kg$^{-1}$day$^{-1}$, a 3$\sigma$ detection of CE$\nu$NS could be achieved with only $\sim 30$~kg-day of exposure. Furthermore, achieving the same sensitivity with a reduced exposure of 14~kg-day would require suppressing the background to below $150\ \mathrm{kg}^{-1}~\mathrm{day}^{-1}$. The full commissioning of the upgraded MINER experiment is planned to take place at the end of 2025, with the potential to significantly improve constraints on CE$\nu$NS cross sections and enable precision measurements.

\section{Acknowledgement}

This work was supported by the U.S. Department of Energy (DOE) under Grant Nos. DE-SC0018981 and DE-SC0017859. The authors gratefully acknowledge seed funding from the Mitchell Institute, which enabled early conceptual development and prototyping efforts, as well as the operating costs of this experiment. We acknowledge the support of Texas A\&M's Nuclear Science Center management and staff for access to the reactor, associated facilities, and technical support/resources. We also acknowledge the support of the Department of Atomic Energy (DAE) and the Department of Science and Technology (DST), Government of India. This work is partly supported through the J.C. Bose Fellowship of the Anusandhan National Research Foundation (ANRF) awarded to B.M. The work of J.L.N. is supported by the Australian Research Council through the ARC center of Excellence for Dark Matter Particle Physics, CE200100008. The authors thank Mr. Sudipta Das for valuable discussions and his contributions to the signal modeling framework. The authors would like to acknowledge the use of the Kanaad HPC cluster facility at SPS, NISER.
\bibliography{references}

\end{document}